\documentclass[12pt]{article}
\def\lesssim{{\
\lower-1.2pt\vbox{\hbox{\rlap{$<$}\lower5pt\vbox{\hbox{$\sim$}}}}\ }} 
\def\gtrsim{{\
\lower-1.2pt\vbox{\hbox{\rlap{$>$}\lower5pt\vbox{\hbox{$\sim$}}}}\ }}

\begin{document}

\begin{titlepage}

\begin{flushright}
    March 1997
\end{flushright}

\vspace{0.5cm}

\begin{center}
    {\Large\bf Are Isocurvature Fluctuations of the M-theory Axion
    Observable?}\\
    \vspace{1.5cm} 
    {\large M.~Kawasaki$^{a,*}$ and
    T.~Yanagida$^{b,\dagger}$}\\
    \vspace{1cm}
    {\it $^{a}$Institute for Cosmic Ray Research, University of Tokyo,
    Tokyo 188, Japan\\
    $^{b}$Department of Physics, University of Tokyo, Tokyo 133, Japan\\
    $^{*}$e-mail:kawasaki@icrr.u-tokyo.ac.jp\\
    $^{\dagger}$e-mail:yanagida@kanquro.phys.s.u-tokyo.ac.jp}
\end{center}

\vspace{2.0cm}

\begin{abstract}
    Banks and Dine have recently shown that the M theory naturally
    accommodates the Peccei-Quinn axion. Since the decay constant
    $F_a$ of the axion is large as $F_a \simeq 10^{15}-10^{16}$GeV,
    the halo axion is hardly detected in coming  axion-search
    experiments. However, we show that isocurvature fluctuations of
    the M-theory axion produced at the inflationary epoch are most
    likely detectable in future satellite experiments on anisotropies
    of the cosmic microwave background radiation.
\end{abstract}

\end{titlepage}
\newpage


If the unification scale ($\sim 10^{16}$GeV) of three known gauge
groups corresponds to compactification scale of extra six dimensional
space, the string theory is in a strong-coupling regime. It has been
argued by Horava and Witten~\cite{Horava} that the strong-coupling
heterotic string theory is M theory whose low-energy limit is well
described by eleven dimensional supergravity. In this M-theory the
fundamental energy scale is the eleven dimensional Planck mass
$M_{11}$ ($\simeq2\times 10^{16}$GeV) rather than the four dimensional
one $M_{4}$ ($\simeq 2\times 10^{18}$GeV)~\cite{Witten}. The four
dimensional Planck mass $M_{4}$ is only an effective parameter at low
energies.

This M-theory description of strong-coupling heterotic string theory
leads to various interesting new phenomenologies. Recently, it has
been pointed out by Banks and Dine~\cite{Banks} that some of string
axions survive at low energies, since the world-sheet instanton
effects are suppressed owing to the large compactification radius in
string tension unit.  Thus, they play a role of the Peccei-Quinn
axion~\cite{Peccei}.

The decay constant $F_a$ of the M-theory axion is expected
as~\cite{Banks}
\begin{equation}
    \label{Fa}
    F_a \simeq 10^{15} - 10^{16} \mathrm{GeV},
\end{equation}
which contradicts the constraint, $10^{10}\mathrm{GeV} \lesssim F_a
\lesssim 10^{12}\mathrm{GeV}$, derived in the standard
cosmology~\cite{Kolb-Turner}. However, this problem may be solved by
late-time entropy production through decays of moduli (or Polonyi)
fields~\cite{KMY,Banks2}. Since the late-time decays produce so many
LSP that the universe is overclosed, the LSP must be
unstable~\cite{KMY2}. This implies rather an interesting case in which
axions are the dark matter in the present universe. Unfortunately, the
halo axions are hardly detected in near future axion-search
experiments, since the decay constant $F_a$ is too large as shown in
eq.(\ref{Fa}).

In this short letter we point out that inflationary universe produces
isocurvature fluctuations of the M-theory axion which are most likely
detectable in future satellite experiments on anisotropies of the
cosmic microwave background radiation (CMBR).

Let us first discuss inflaton sector. Since the fundamental energy
scale is $M_{11} \sim 2\times 10^{16} \mathrm{GeV}$ in the M
theory~\cite{Witten,Banks}, it is natural to consider that the
inflaton takes a value of the order of $M_{11}$ at the beginning of
the universe. Therefore, the most natural inflation model seems to be
the hybrid inflation model~\cite{Linde}. In the hybrid inflation model
the inflation occurs with a constant vacuum energy density which is
quickly eaten by another field when the inflaton reaches a critical
value. The first guess of the vacuum energy density may be $M_{11}^4$.
However, it is too large to have sufficiently long inflationary epoch.
This is because the initial value of the inflaton is about
the same as the critical value at which the inflation ends. The next
choice is $M_{11}^6/M_{4}^2 \sim (2\times 10^{15} \mathrm{GeV})^4$ as
argued in ref.~\cite{Banks2}. This vacuum energy density is also suggested by
a recent analysis~\cite{Linde2} in a hybrid
inflation model.\footnote{
Linde and Riotto~\cite{Linde2} have recently proposed a hybrid
inflation model based on supergravity. From COBE data they have derived 
the vacuum energy density $V \sim (2\times 10^{15}\mathrm{GeV})^4$.}
Thus, we assume  hereafter the vacuum energy density
\begin{equation}
    \label{energy}
    V \sim (2\times 10^{15}\mathrm{GeV})^4
\end{equation}
during the inflation, which leads to the Hubble
constant
\begin{equation}
    \label{hubble}
    H_{\mathrm{inf}} \simeq \frac{\sqrt{V}}{\sqrt{3}M_4}
    \sim 10^{12}\mathrm{GeV}.
\end{equation}

We are now at the point to estimate isocurvature fluctuations of the
M-theory axion. It is known~\cite{Linde3} that in the deSitter
universe massless fields $\varphi$ have quantum fluctuations $\delta
\varphi$ which are given by 
\begin{equation}
    \label{delta-phi}
    \delta \varphi \simeq \frac{H_{\mathrm inf}}{2\pi}.
\end{equation}
Since the axion does not have potential energy during inflation, its
fluctuations do not give any contribution to those of the total energy
density of the universe. Thus the axion fluctuations are of
isocurvature type. The isocurvature fluctuations of the dark matter
density (provided that the axion is the dark matter) are given by
\begin{equation}
    \label{iso}
    \left(\frac{\delta\rho_a}{\rho_a}\right)_{\mathrm{iso}}
    \simeq \frac{H_{\mathrm{inf}}}{\pi F_a}
    \simeq (3 - 0.3)\times 10^{-4}.
\end{equation}
Here we have used eqs.(\ref{Fa}) and (\ref{hubble}). On the other
hand, the inflaton itself produces adiabatic fluctuations as
usual. Thus we have a mixture model of adiabatic and isocurvature
fluctuations of the dark matter (axion) density.

The observation of the anisotropies of CMBR by COBE~\cite{COBE} gives
a constraint on the isocurvature fluctuations as
\begin{equation}
    \label{COBE-iso}
    \left(\frac{\delta \rho}{\rho}\right)_{\mathrm{iso}}
    \lesssim 2\times 10^{-5}.
\end{equation}
The prediction in eq.(\ref{iso}) already violates this
constraint. However, we should not take eq.(\ref{iso}) at the face
value, since our estimation of $H_{\mathrm{inf}}$
and $F_a$ may contain various ambiguities. The above analysis, nevertheless,
shows that the isocurvature fluctuations of the M-theory axion most
likely give a non-negligible contribution to CMBR.\footnote{
Interestingly, it has been pointed out~\cite{KSY} that if
isocurvature fluctuations are comparably mixed with adiabatic ones,
the matter power spectrum normalized by COBE data give a better fit to
the data of observations of the large scale structure of the universe
than in the case of pure adiabatic fluctuations. In the standard cold
dark matter scenario with pure adiabatic fluctuations, the
density fluctuations at scales of galaxies and clusters are too large
if the power spectrum $P(k)$ is normalized by COBE data.  However,
since isocurvature fluctuations give a six times larger contribution
to CMBR anisotropies at COBE scales, the mixture of isocurvature
fluctuations decreases the amplitude of the matter fluctuations at
scales of galaxies and clusters, which leads to a better fit to the
observations. }

If $(\delta \rho_a/\rho_a)_\mathrm{iso} \simeq (1-2)\times 10^{-5}$,
anisotropies of CMBR induced by isocurvature fluctuations can be
distinguished from those produced by pure adiabatic
fluctuations~\cite{KSY} because the shapes of power spectrum of CMBR
anisotropies are quite different from each other at small angular
scales.  For example, there will be no large Doppler peak in the
spectrum of CMBR as pointed out in ref.~\cite{KSY}, while the pure
adiabatic fluctuation models predict a significant peak at degree
scales. Since direct searches for the M-theory axion seem impossible,
the future satellite experiments on CMBR anisotropies is crucial to
test the M-theory axion hypothesis.

\vspace{0.5cm}
\noindent
\textbf{Acknowledgment}

One of the authors (T.Y.) thanks M. Hashimoto and K.-I. Izawa for
useful discussions.

\end{document}